\documentclass{ws-procs975x65}
\usepackage{graphicx}
\usepackage{multicol}

\def\beq{\begin{equation}}
\def\eeq{\end{equation}}

\begin{document}

\title{Tests of GR with INPOP15a planetary ephemerides: estimations of possible supplementary advances of perihelia for Mercury and Saturn.}

\author{Agn\`es Fienga$^*$}
\address{G\'eoAzur, CNRS-UMR7329, Observatoire de la C\^ote d'Azur, \\
Universit\'e Nice Sophia Antipolis, Valbonne, 06560, France\\
$^*$E-mail: fienga@oca.eu}

\author{Jacques Laskar$^*$, Herv\'e Manche and Mickael Gastineau} 
\address{IMCCE, CNRS-UMR8028, Observatoire de Paris,\\
Universit\'e Pierre et Marie Curie, Paris, 75014, France\\
$^*$E-mail: laskar@obspm.fr}

\begin{abstract}
Planetary ephemerides are a good tool for studying general relativity at the scale of our solar system. We present here new evaluations of advances of perihelia for Mercury and Saturn as well as new boundaries for possible values for PPN parameters. 
\end{abstract}


\bodymatter


\section{New planetary ephemerides INPOP15a}

On the basis of the INPOP planetary ephemerides construction, the INPOP15a was built in adding supplementary range tracking data obtained from the analysis of the MESSENGER spacecraft from 2011 to 2014 and in including the new JPL datasets obtained after the new analysis of Cassini tracking data obtained from 2004 to 2014 (Ref.~\refcite{Folkner2014}). 
As described by Ref.~\refcite{Folkner2014}, this new analysis had corrected a mis-estimation of the Saturn positions based on an insufficient convergence of the previous fitting procedure. Furthermore in comparison to the previous Cassini data sets used in DE423 and in INPOP13c, the interval covered by the observations increases.
Differences between INPOP15a and other planetary ephemerides built with almost the same data sampling but including the differences in the the dynamical modelings and in the fitting procedures will give a realistic estimation of the present uncertainties in planetary orbits.
As described in Ref.~\refcite{Fienga2015} such differences can reach up to 50$\%$ in the postfit residuals between INPOP13c residuals and DE430 residuals when for INPOP15a the differences between residuals are only at the maximum of about 35$\%$.
As for Ref.~\refcite{Fienga2015}, we will consider this latest condition $ \Delta (O-C) < 35\%$ as a acceptable threshold for considering an ephemeris significantly different from INPOP or DE430.

\section{Supplementary advances of perihelia}

Following the method described in Ref.~\refcite{Fienga2011}, supplementary advances of perihelia were added to the ephemerides. Full adjustement of the new ephemerides including supplementary advances were operated. Only ephemerides with differences between their postfit residuals and INPOP15a residuals such as  $\Delta (O-C) < 35\%$ (criteria 1) are seen as acceptable. 
In Ref.~\refcite{Fienga2011} the advances of perihelia were obtained by considering $ \Delta (O-C) < 5\%$ (criteria 2). So in order to make comparisons, Table \ref{tab1} gives the intervals for possible advances of perihelia for the criteria 1 in Column 2, the criteria 2 in Column 3 and the mean of the values obtained with the two in Column 4. For comparison, the values obtained with INPOP10a (Ref.~\refcite{Fienga2011}) and those published by Ref.~\refcite{Pitjeva2013} are also given in the Column 5 and 6 respectively.
Finally as the orbits of Mercury and Saturn have been drastically improved in comparison to INPOP10a and INPOP13c, we present the advances of perihelia for these two planets.

\begin{table}
\caption{Supplementary advances of perihelia $\varpi_{sup}$ obtained with INPOP15a, and INPOP10a(Ref.4) in one hand and Ref.3 in the other hand.}
\begin{tabular}{c c c c c c c}
\hline
& \multicolumn{3}{c}{INPOP15a} & INPOP10a& P13\\
$\varpi_{sup}$ &  criteria 1  & criteria 2  & (C1+C2)/2 & \\
mas.cy$^{-1}$ &  (C1) & (C2) & &  \\
\hline
Mercury & (0.0 $\pm$ 3.1) & (0.0 $\pm$ 1.05)  & (0.0 $\pm$ 2.075) &  (1.2 $\pm$ 1.6)&  (-2 $\pm$ 3)\\
\\
Saturn & (1.2 $\pm$ 5.0) & (0.05 $\pm$ 0.20) &  (0.625 $\pm$ 2.6) & (0.15 $\pm$ 0.65) &  (-0.32 $\pm$ 0.47)\\
\hline
\end{tabular}
\label{tab1}
\end{table}

No departure from 0 are noticed for Mercury and Saturn, inducing no possible violation of GR at this level of accuracy. One can noticed that the INPOP10a estimations are compatible with the INPOP15a. However, The INPOP15a Saturn possible advance of perihelia is significantly larger than the INPOP10a one. This can be explained by the new analysis of the Saturn Cassini data over a longer time span (1 year for INPOP10a to 10 years for INPOP15a) and the removal of some systematic trend in the raw tracking data of the Cassini spacecraft by the JPL team (Ref.~\refcite{Folkner2014}). 
The values obtained by Ref.~\refcite{Pitjeva2013} follow the same pattern as INPOP10a values as they were obtained in using about the same sample of data. It will be interesting to compare the latest INPOP values with the new estimations done by Pitjeva in using the new Cassini data.
In terms of comparison to alternative theories of gravitation and in particular to MOND (Ref.~\refcite{blanchet11}), the new estimated values are still competitive for selecting possible MOND functions. Following the Ref.~\refcite{blanchet11} notations, only $\mu_{20(y)}$ is indeed the only possible functions regarding the impact of the quadrupole term $Q_{2}$ on the advance of planet perihelia. 

\end{document}